\documentclass[12pt,showpacs,preprintnumbers,amsmath,aps,amssymb,epsfig]{revtex4}
\usepackage{graphicx}
\usepackage{dcolumn}
\usepackage{bm}

\begin{document}

\title{Iso-spectral Potentials and Inflationary quantum Cosmology}
\author{A. Garc\'{\i}a }
 \email{alberto@fisica.ugto.mx}
\author{W. Guzm\'an}
 \email{wguzman@fisica.ugto.mx}
 \author{M. Sabido}
\email{msabido@fisica.ugto.mx}
\author{J. Socorro}
\email{socorro@fisica.ugto.mx}
\affiliation{Instituto de F\'{\i}sica de la Universidad de Guanajuato,\\
 A.P. E-143, C.P. 37150, Le\'on, Guanajuato, M\'exico}%

\date{\today}

\begin{abstract}
 Using the factorization approach of quantum mechanics, we obtain a family of 
 isospectral scalar potentials for power law inflationary cosmology. The construction is based 
 on a scattering Wheeler-DeWitt solution. These iso-potentials have new features, they give a mechanism to end inflation, as well as the possibility to have new inflationary epochs. The procedure can be extended to other cosmological models.
\end{abstract}

\maketitle

\section{Introduction}
One of the most active areas of research nowadays is inflationary cosmology, this theoretical framework
solves many classical problems of  Standard Big Bang Cosmology.
 Recently, observations have confirmed its predictions of a flat 
universe with a nearly scale invariant perturbations spectrum. The idea is to introduce a scalar field (spin-0 boson) and a scalar potential $V(\phi)$ which encodes in itself the (non-gravitational) self-interactions among the scalar particles.
This type of models, have also been used within the so
called canonical Quantum Cosmology (QC) formalism, which deals with the early epoch of the cosmos. Scalar fields act as
matter sources, and then play an important role in determining the
evolution of the early universe, where the quantum fluctuations are the seeds for structure formation.

Moreover, these models have appeared in String Theory, in particular in connection to the so called string theory
landscape as well as in the study of tachyon dynamics. For the String Theory landscape
 (Kobakhidze and Mersini-Houghton, 2004; Douglas, 2003; Susskind, 2003), the scalar
potential $\rm V(\phi)$ is usually thought as having many valleys, which
represent the different vacua solutions, the hope is that the
statistics of these vacua could explain, for example, the smallness of
the cosmological constant (the simplest candidate for dark energy). For tachyon dynamics in the unstable D-brane scenario, the scalar potential for
the tachyon effective action around the minimum of the potential has
the form $\rm V(\phi)=e^{-\alpha \phi/2}$ (Sen, 2002,2003). 
This leads to the study of tachyon driven
cosmology (Gorini 2004; Garcia-Compean, Garc\'{\i}a-Jimenez, Obreg\'on 
	and Ram\'{\i}rez, 2005). 

On another front, the study of eigenvalue problems associated with second-order differential
operators found a renewed interest with the application of the factorization
technique and its generalizations (Cooper, Khare and Sukhatme, 1995; Fern\'andez, 1984; 
Mielnik, 1984; Nieto, 1984; Gelfand and Levitan, 1955).
SUSY-QM may be considered an equivalent formulation of the Darboux 
transformation method, which is  well-known in mathematics from the original 
paper of Darboux (Darboux, 1982; Ince, 1926). 
An essential ingredient is a differential
operator (Bagrov and  Samsonov, 1995) which intertwines two hamiltonians and relates their
eigenfunctions. When this approach is applied in quantum theory it allows 
to generate a family of exactly solvable local potential starting
with a given exactly solvable one (Cooper, Khare and Sukhatme, 1995). 
In nonrelativistic one-dimensional supersymmetric quan\-tum mechanics, the
factorization technique was applied to the $\rm q=0$ factor ordered WDW 
equation corresponding to the FRW cosmological models without matter field 
( Rosu and Socorro, 1998), where a one-parameter class of strictly isospectral cosmological
FRW solutions was exactly found, representing the wave functions of the 
universe for that case, also iso-spectral solutions for a  one-parameter family of 
closed, radiation-filled FRW quantum universe, and
 for a perfect fluid with barotropic state equation and cosmological constant term, 
for any factor ordering were found (Rosu and Socorro, 1996; Socorro, Reyes and Gelbert, 2003). 
In this formalism, the family of iso-potentials and
wave functions are build with respect to a parameter $\gamma_i$ for which we
choose the domain $[0,\infty ]$. The shape of the wave function in the corresponding
coordinates is obtained via the ``evolution" of the iso-wave function when this parameter 
tends to $\infty$.

The main purpose of this paper is to apply the Darboux transformation method to  
obtain a family of  iso-potentials, to the potential $V(\phi)=e^{-\alpha\phi/2}$, which appears in inflationary cosmology. To reach this goal, we shall make use of the strictly 
isospectral scheme based on the general Riccati solution
(Cooper, Khare and Sukhatme, 1995; Fern\'andez, 1984; 
Mielnik, 1984; Nieto, 1984), which is also known as the double Darboux method.
This scheme has been applied
from classical and quantum physics (Mielnik,1984)  to relativistic models
(Samsonov and  Suzko, 2003). This  technique has been known for a decade in 
one-dimensional supersymmetric quantum mechanics (SUSY-QM) and usually 
requires 
nodeless, normalizable states of a
Schr\"odinger-like equation. However, Pappademos, Sukhatme, and Pagnamenta (1993)
 showed that the strictly isospectral construction
can also be performed on non-normalizable states. The resulting potentials 
have interesting features, in particular they solve the one problem the exponential 
potential has, the lack of a mechanism to end inflation.

This work is   organized as follows. In section II we present 
the classical action with the corresponding contributions, this action includes a 
gravitational part $\rm S_g$, and $\rm S_\phi$ for the scalar field; also  we present
the standard quantum scheme with the quantum solution, which plays an
important role in the isospectral solutions. In section III we review the factorization approach of (Susy-QM) and apply it to the inflationary model. Finally section IV
is devoted to conclusions and outlook.

\section{The standard quantum scheme}

We start with the line element for a homogeneous and isotropic
universe, the so called Friedmann-Robertson-Walker (FRW) metric, in the form
\begin{equation}
\rm ds^2= -N^2(t) dt^2 + e^{2\alpha(t)}\left[\frac{dr^2}{1-k r^2}
+r^2 (d\theta^2 + sin^2 \theta d\varphi^2) \right] \, ,
\end{equation}
where $\rm a(t)=e^{\alpha(t)}$ is the scale factor, $\rm N(t)$ is the lapse function, and
$\rm \kappa$ is the curvature constant that takes the values $\rm 0,+1,-1$, which
correspond to a flat, closed or open universe, respectively. 

The effective action we will be working, is 
( W. Guzm\'an, Sabido, Socorro and Arturo Urena L\'opez, 2005)

\begin{equation}
\rm S_{tot}= S_g +S_{\phi} =\int  dx^4 \sqrt{-g} \left[ R + 
 \frac{1}{2}g^{\mu \nu}\partial_\mu \phi \partial_\nu \phi+ 
 V_0 e^{-\frac{\lambda}{\sqrt{12}}\phi}
  \right] \, ,
\label{accion1}
\end{equation}
$\rm \phi$ is a scalar field endowed with a scalar potential 
 $\rm V(\phi) = V_0 e^{-\frac{\lambda}{\sqrt{12}}\phi}$.

 The Lagrangian
for a  FRW cosmological model is
\begin{equation}
\rm {\cal L}=e^{3\alpha} \left[6 \frac{ \dot{\alpha}^2}{N} -
\frac{1}{2} \frac{\dot{\phi^2}}{N} + 
N \left( V(\phi) - 6 \kappa e^{-2\alpha} \right) \right]\, ,
\label{lagra}
\end{equation}
At this point, we consider a flat universe ($\rm \kappa=0$) 
\begin{equation}
\rm {\cal L}=e^{3\alpha} \left[6 \frac{ \dot{\alpha}^2}{N} -
\frac{1}{2} \frac{\dot{\phi^2}}{N} + 
N  V(\phi) \right]\, ,
\end{equation}

The canonical momenta are found to be
\begin{subequations}
\begin{eqnarray}
\rm \Pi_\alpha&=&\rm \frac {\partial {\cal L}}{\partial \dot{\alpha} }=
12 e^{3\alpha}\frac{\dot{\alpha}}{N} \, ,
\qquad \dot{\alpha}=\frac{N}{12} e^{-3\alpha} \Pi_\alpha \, , \label{pa}\\
\rm \Pi_\phi&=&\rm \frac {\partial {\cal L}}{\partial \dot{\phi} }=
- e^{3\alpha}\frac{\dot{\phi}}{N} 
\, , \qquad\dot{\phi}=-N e^{-3\alpha} \Pi_\phi \, . \label{pp}
\end{eqnarray}
\end{subequations}
We are now in position to write the corresponding canonical Hamiltonian (Ryan, 1972)
\begin{equation}
\rm {\cal L}_{canonical}= \Pi_\alpha \dot \alpha + \Pi_\phi \dot \phi - N {\cal H},
\label{canonical}
\end{equation}
where $\cal H$ is the classical Hamiltonian function, having the following structure
\begin{equation}
\rm {\cal H}=\frac{1}{24} e^{-3\alpha} \left[
\Pi_\alpha^2 - 12\Pi_\phi^2 - 24  e^{6 \alpha}  V(\phi)
\right] \, , \label{uno}
\end{equation}
and performing the variation of (\ref{canonical}) with respect to N,
$\rm {\partial {\cal L}}/{\partial N}= 0$, implies the well-known
result $\rm {\cal H}=0$.
The Wheeler-DeWitt (WDW) equation for this model is achieved by replacing 
$\rm \Pi_{q^\mu}$
by $\rm -i \partial_{q^\mu}$ in Eq.~(\ref{uno}); here $\rm q^\mu=(\alpha, \phi)$.

   Under a particular factor ordering the WDW reads
\begin{equation}
 \rm \hat H =   \frac{e^{-3\alpha}}{24}\left[-\frac{\partial^2}{\partial\alpha^2}
  +12\frac{\partial^2}{\partial\phi^2}  -24e^{6\alpha}{ V}(\phi)
 \right] \Psi= 0. \label{WDW1} 
 \end{equation}
or
\begin{equation}\rm 
\Box \, \Psi  - 24 e^{6\alpha} \tilde V(\phi) \Psi =0 \, , 
\label {WDW}
\end{equation}
with $\rm \tilde \phi =\frac{\phi}{\sqrt{12}}$,
 $\Psi$ is called the wave function of the uni\-ver\-se, $\rm \Box \equiv
  -\partial^2_\alpha + \partial^2_{\tilde \phi}$ is the two dimensional
  d'Alambertian operator in the $q^\mu$ coordinates. 
From now on we fix the potential to $V(\phi)=e^{-\lambda\tilde\phi}$. Applying the factorization method in these variables is technically cumbersome, this can be simplified if we make the following coordinates transformation,
\begin{equation}
\rm x = 6 \alpha - \lambda \tilde \phi, \qquad y = \alpha - \frac{6}{\lambda} \tilde \phi,
\label{trans}
\end{equation}
the WDW equation ({\ref{WDW}) takes the form
\begin{equation}
\rm  \frac{\partial^2 \Psi}{\partial x^2}
 -\frac{1}{\lambda^2} \frac{\partial^2 \Psi}{\partial y^2} - \frac{24 V_0}{\lambda^2 - 36 }
 e^x \Psi=0
\label{modified}
\end{equation}
and by separation variables, $\rm \Psi=X(x) Y(\tilde y)$ with 
$\rm \tilde y =\lambda y$, we obtain the set
of differential equation for the functions X and Y
\begin{eqnarray}
\rm \frac{d^2 X}{dx^2} + \left(- \beta e^x + \frac{\eta^2}{4}  \right) X&=&0,
\nonumber\\
\rm \frac{d^2 Y}{d \tilde y^2}+ \frac{\eta^2}{4}  Y &=&0,
\label{hams}
\end{eqnarray}
where 
\begin{equation}
\rm \beta= \frac{24 V_0}{\lambda^2 -36},
\label{beta}
\end{equation}
we choose for simplicity $\frac{\eta^2}{4}$ as a separation constant. The
solutions for these equations are
\begin{eqnarray}
\rm X(x) &=& \rm Z_{\pm i\eta} \left( \pm 2i\sqrt{\beta} e^{x/2} \right) ,  \label{u1}\\
\rm Y(\tilde y) &=& \rm A_0 e^{i\frac{\eta \lambda}{2} \tilde y} + A_1 e^{-i \frac{\eta \lambda}{2}\tilde y},
\label{u2},
\end{eqnarray}
with $\rm Z_{\pm i\eta}$   are generic Bessel Functions
 with pure imaginary order, the wave function
is 
\begin{equation}\rm 
\Psi_\eta(x,\tilde y)= e^{\pm i \frac{\eta \lambda}{2}\tilde y} 
Z_{i\eta}\left( \pm 2i\sqrt{\beta} e^{x/2} \right). 
\label{function}
\end{equation}

 Since these solutions have the dependence in the parameter $\eta$,  the general
solution can be put as
\begin{equation}
\Psi_{gen} = \int G(\eta) \Psi_\eta d\eta,
\end{equation}
where $G(\eta)$ represents a weigh function.

The selection of the value of $\lambda$ in (\ref{beta}), gives the structure for the
 $\rm Z_{\pm i\eta}$, for $\lambda>6$, we have the modified Bessel function,
 and for $0<\lambda<6$, $\rm Z_{\pm i\eta}$ become the ordinary Bessel function. 
 We are now in a position to form a normalizable {\it Gaussian state} as a superposition 
 of the eigenfunctions (\ref{function}). With this in mind a {\it wave packet} can be 
 constructed (Kiefer,1988,1990).
 
 We can have different solutions that depend on the value of the parameter $\lambda$. 
For $\lambda>6$, the {\it wave packet} can be constructed 
using the modified Bessel function (see Gradshteyn and  Ryzhik, 1980 ) 
\begin{equation}\rm 
\Psi (x, \tilde y) = \int_0^\infty  \cos \left( \frac{\eta \lambda}{2}\tilde y \right)
 K_{i\eta}\left( \pm 2\sqrt{\beta} e^{x/2} \right) d\eta= \frac{\pi}{2} 
 Exp\left[-2\sqrt{\beta} e^{x/2} 
 cosh\left( \frac{\lambda}{2}\tilde y \right)  \right].
\end{equation}
In the range $0<\lambda<6$ ( which includes the inflation), we can also construct a 
{\it wave packet}, but this time using the ordinary  Bessel functions 
\begin{equation}\rm 
\Psi (x,\tilde y) = \int_{-\infty}^\infty \frac{e^{\pi x/2}}{sinh(\pi x)} 
\cos \left( \frac{\eta \lambda}{2}\tilde y \right)
 J_{i\eta}\left( \pm 2\sqrt{\beta} e^{x/2} \right) d\eta= -i 
 Exp\left[i 2\sqrt{\beta} e^{x/2} 
 cosh\left( \frac{\lambda}{2}\tilde y \right)  \right].
\end{equation}
We have been using the new variables $x$ and $\tilde y$. Let us now extract 
some information from the semiclassical behavior as a check of our quantum model.
The classical solutions can be obtained using the  semiclassical analysis 
(WKB-like method). 
For this, one considers the ansatz on the wave function
\begin{equation}
\rm \Psi(\alpha,\phi)= e^{-S},
\end{equation}
and the conditions 
\begin{equation}\rm
\left( \frac{\partial S}{\partial a}\right)^2 \gg \left| 
\frac{\partial^2S}{\partial a^2}\right| ,
\qquad \quad
\left( \frac{\partial S}{\partial \phi }\right)^2 \gg \left| 
\frac{\partial^2 S}{\partial \tilde \phi^2}\right|,  
\label{condiciones}
\end{equation}
from this, the Einstein-Hamilton-Jacobi equation (EHJ) is obtained,
and Eq. (\ref{WDW}) reads
  \begin{equation} \rm 
 \left(\frac{\partial S}{\partial \alpha}  \right)^2 -
 \left(\frac{\partial S}{\partial \tilde \phi}  \right)^2 
 - 24 V_0 e^{6\alpha -\lambda \phi}= 0, \label {hj1}
\end{equation}

The same equation is recovered directly when we introduce the transformation 
on the canonical momentas 
\begin{equation}
\rm \Pi_{q^\mu}\rightarrow \frac{\partial S}{\partial q^\mu}, \label{mome}
\end{equation}
in Eq. (\ref{uno}), in the new coordinates $x$ and $y$, this equation takes the form
\begin{equation}
\rm  \left( \frac{\partial S}{\partial x} \right)^2
-  \left( \frac{\partial S}{\partial \tilde y} \right)^2 -\beta e^{-x} =0.
\end {equation}
and choosing $\rm S=S_x S_y$ we obtain the following solutions
\begin{eqnarray}
\rm S_x &= &\rm \pm \frac{2}{\sqrt{\beta}\mu} e^{-x/2}, \nonumber\\
\rm S_{\tilde y} &=& \rm \mu =cte.
\end{eqnarray}
so we get, 
\begin{equation}
\rm S(x,\tilde y)= \pm 2\sqrt{\beta} e^{-x/2}.
\end{equation}
Equation (\ref{mome}) in the new variables become
\begin{eqnarray}
\rm \Pi_\alpha&=& \rm \pm \frac{6}{\sqrt{\beta}} e^{-x/2}=\rm \pm \frac{6}{\sqrt{\beta}} 
e^{3\alpha-\frac{\lambda}{2}\phi},\nonumber\\
\rm \Pi_\phi &=& \rm \mp \frac{\lambda}{\sqrt{\beta}} e^{-x/2}=\rm \pm \frac{\lambda}{\sqrt{\beta}} 
e^{3\alpha-\frac{\lambda}{2}\phi} ,
\label{mome1}
\end{eqnarray}

The classical behaviour is considered solving the relations between 
(\ref{mome1})  and Eqs. (\ref{pa},\ref{pp}), obtaining the relation for the
the scale factor and the scalar field 
\begin{equation}
\rm a= a_0 e^{\frac{1}{2\lambda} \tilde \phi},
\end{equation}
the corresponding time behaviour ($\rm d\tau=Ndt$),
\begin{eqnarray}
\rm a &=& \rm a_0 \tau^{\frac{2}{\lambda^2}}, 
\label{scalefactor}\nonumber\\
\rm \tilde \phi &= &\rm  \frac{2}{\lambda} 
Ln\left(\frac{ \lambda^2}{4\sqrt{3\beta}}  \tau + \tilde \phi_0 \right),
\end{eqnarray}
having an  inflationary scenario for the scale factor, with increasing power law 
when  $\rm \lambda< \sqrt{2}$. This is the result from the classical approach 
(Chimento and  Jakubi,1996). Unfortunately this potential has a serious drawback, 
once we fix the value of $\lambda$ to for inflation, there is no way to end it.

\section{Supersymmetric factorization scheme}

As already mentioned, the goal of this paper is to use the factorization 
approach of supersymmetric quantum mechanics to obtain the family of isoespectral 
potentials to $V(\phi)=e^{-\lambda\tilde\phi}$, an  see if these potentials lead 
to new physics.

If a quantum system is characterized by Hamiltonian $H$ with eigenvalues $\nu$, 
we may ask if there exists another Hamiltonian (i.e. different potential) which has 
the same spectrum. This is an interesting question because  we may not be able to  
distinguish one system from the other by making them physically equivalent. With 
this in mind, we will begin by reviewing the factorization approach.

Lets start with the hamiltonian  that appears in the study of scalar fields interacting
with gravitation Eq.(\ref{modified}, \ref{hams}), written in generic form as
\begin{eqnarray}
\rm -\frac{d^2P_i}{dq_i^2} +    V_i(q_i)P_i &=& \rm  E_i P_i, \quad P_i=(X,Y), \quad q_i=(x,y), \nonumber\\
\rm E_i & =& \left\{ 
\begin{tabular}{l}
$ \frac{\nu^2}{4\mu^2 }  $\\
$ \frac{B^2 \nu^2}{\mu^2 }   $
\end{tabular} \right.,\nonumber\\
\rm V_i & =& \left\{ 
\begin{tabular}{l}
$-\frac{6 }{\mu^2} e^x $\\
$0$
\end{tabular} \right..
\label{like}
\end{eqnarray}
It is easy to show that the first  order  differential operators
\begin{eqnarray}
{\cal A}_i^+ &=& {\rm  -  \frac{d}{d q^i} + W_i(q^i)}, \label{a+}\\
{\cal A}_i^- &=& {\rm   \frac{d}{d q^i} + W_i(q^i)}, \label{a-}
\end{eqnarray}
factorize the hamiltonian, l.h.s.of Eq. (\ref{like})
(here $W_i$ plays the role of a superpotential function) as
\begin{equation}
\rm H^+_i-E_i= {\cal A}_i^+ {\cal A}_i^- \,  \qquad i=1,2.
\label {h+}
\end{equation}
The potential term $\rm V_i(q^\mu)$ is related to the superpotential function
$\rm W_i(q^\mu)$ via the Riccati equation \footnote{ we shall call 
$\rm V^+_i$ to $\rm V_i$ in (\ref{u1}) and (\ref{u2}).}
\begin{equation}
\rm V_i(q^i) - E_i= W^2_i -\frac{dW_i}{dq^i}, \qquad i=1,2
\label{ricatti}
\end{equation}
Making the transformation 
\begin{equation}
\rm W_i= - \frac{u^{\prime}_i}{u_i}, \quad i=1,2
\label{superpotential}
\end{equation}
where the $\prime$ means $\rm \frac{d}{dq^i}$,  
(\ref{ricatti}) is transformed into the original hamiltonians applied to the
functions $\rm u_i$ that correspond to the solutions (\ref{u1},\ref{u2}), this implies
that once we have a solution to the original Schr\"odinger like equation (\ref{like})
the superpotential function can be constructed. In this  factorization 
scheme, $\rm V^-_i$ is 
the partner superpotential of  $\rm V^+_i$, and can be calculated by performing
the product
\begin{equation}
\rm H^-_i -E_i = {\cal A}_i^- {\cal A}_i^+ , \qquad H^-f_i(q^i)=E_i f_i(q^i),
\label{h-}
\end{equation}
where $\rm f_i$ is the wave function related to the hamiltonian $\rm H^-_i$.
Then, the isospectral potential to $\rm V^+_i(q^i)$ is
\begin{equation}
\rm V^-_i(q^i)- E_i= W^2_i +  W^{\prime}_i.
\label{-}
\end{equation}
Using (\ref{ricatti}), we get the functional relation between $\rm V^-_i$ and $\rm V^+_i$, being
\begin{equation}
\rm V^-_i(q^i)= V^+_i(q^i)+ 2  W^{\prime}_i .
\label{iso} 
\end{equation}
However, this is not the most general
solution as  will be shown in the following section.

\subsection{General Solution}
The general solution to the Ricatti equation (\ref{iso}) is found from 
(Cooper, Khare and Sukhatme, 1995; Mielnik, 1984)
\begin{equation}
\rm V^-_i(q^i)-E_i=  \hat W^2_i + \hat W^\prime_i \equiv
 W^2_i +  W^{\prime}_i,
\label{general-potential}
\end{equation}
which by choosing 
\begin{equation}
\rm \hat W_i \equiv W_i + \frac{1}{y_i},
\label{general}
\end{equation}
leads to a Bernoulli equation for $\rm y_i$,
\begin{equation}
\rm y^{\prime}_i - 2 W_i \,y_i=1, 
\end{equation}
whose solution is
\begin{equation}
\rm y_i (q^i)=\frac{I_i+ \gamma_i }{u_i^2},
\label{eq42}
\end{equation}
where $\rm I_i(q^i)= \int_0^{q^i} u_i^2 dx$ and $\gamma_i$ is the factorization 
parameter ($\gamma_i$
plays the role of a {\it " time-like parameter"} in the evolution of the isospectral
wavefunction).

By using Eq.(\ref{eq42}), we may write Eq.(\ref{general}) as
\begin{equation}
\rm \hat W_i(q^i)=W_i + \frac{u_i^2}{I_i + \gamma_i},
\end{equation}
and, the entire family of bosonic potentials can be constructed from 
\begin{equation}
\rm \hat V^+_i(q^i, \gamma_i)- E_i= \hat W^2_i(q^i,\gamma_i) 
- \hat W^\prime_i(q^i,\gamma_i),
\end{equation}
or
\begin{eqnarray}
{\rm \hat V^+_i(q^i,\gamma_i)}&=&{\rm  V^-_i - 2 \hat W^\prime_i }\nonumber\\
&=& {\rm V^+_i(q^i) - 4\frac{ u_i u_i^\prime}{I_i 
+\gamma_i} + 2\frac{ u_i^4}{(I_i + \gamma_i)^2} },
\label{iso-pote}
\end{eqnarray}

\begin{figure}
\includegraphics[width=8cm]{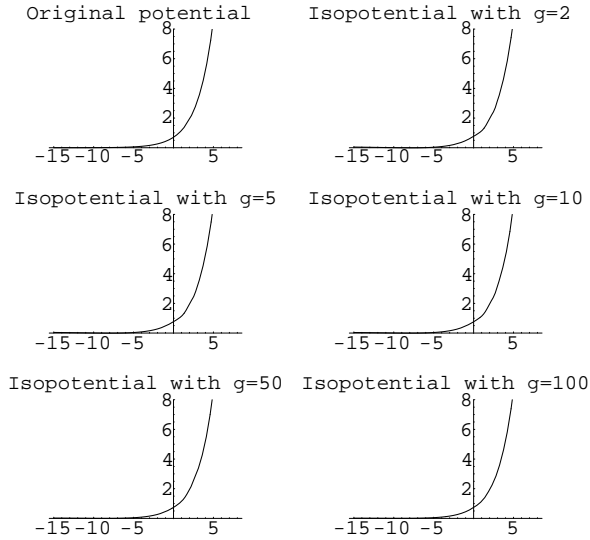}
\caption{\label{fig:esc1} Plot is of the original potential and the iso-potentials. At large scale, the original potential and the iso-potentials do not differ much. }
\end{figure}

\begin{figure}
\includegraphics[width=8cm]{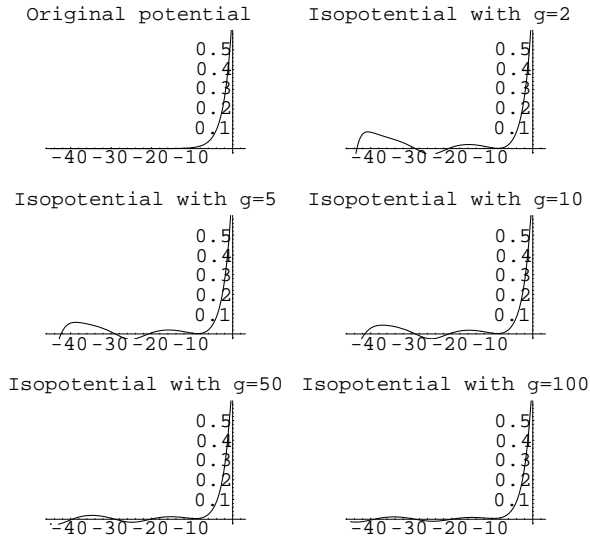}
\caption{\label{fig:esc2} At small scale the iso-potential have new structure for small values in the parameter $\gamma$, giving rise to a mechanism to end inflation that is not present in the original potential.}
\end{figure}
\begin{figure}
\includegraphics[width=8cm]{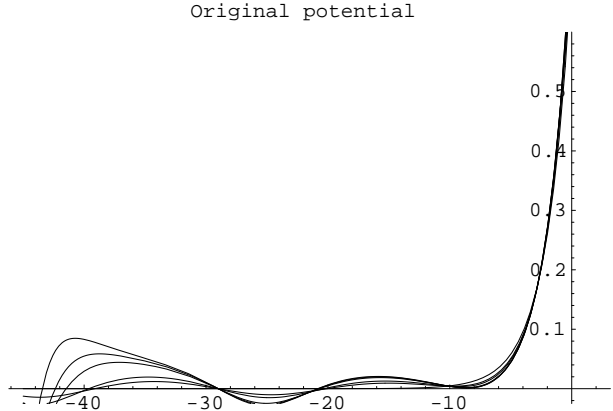}
\caption{\label{fig:esc3} The corresponding changes in the structure for the isopotential,
mainly for small values in the $\gamma$ parameter. }
\end{figure}
finally
\begin{equation}
\rm \hat u_i \equiv  g(\gamma_i) \frac{ u_i}{ I_i + \gamma_i},
\label{solu}
\end{equation}
is the isospectral solution of the Schr\"odinger equations (\ref{u1},\ref{u2})
for the new 
family of potentials (\ref{iso-pote}), with the condition on the function 
$\rm g(\gamma_i)= \sqrt{\gamma_i(\gamma_i+1)}$, though in the limit
\begin{equation}
\rm \gamma_i \to  \infty \qquad g(\gamma_i) \rightarrow \gamma_i \qquad and 
\quad \hat u_i \rightarrow u_i.
\end{equation}
Considering the particular case for $\lambda<\sqrt{2}$ (which correspond to inflation), 
we plot the solutions in the variable $\rm x = 6 \alpha - \lambda \tilde \phi$ for 
the isospectral potential
(\ref{iso-pote}) (figures 1, 2 and 3) and the corresponding wavefunctions (\ref{solu})
(figures 4 and 5). 
The total WDW isospectral wave function has the following form
\begin{equation}
\rm \Psi_{iso}(x,\tilde y;\gamma_1,\gamma_2)=  \frac{
\left[J_{i\eta} \left( \pm 2\sqrt{\beta} e^{x/2} \right) 
+ J_{-i\eta} \left( \pm 2\sqrt{\beta} e^{x/2}  \right) \right]}{I_1 + \gamma_1} 
\times 
 \frac{\left[a_0 e^{i\frac{\eta \lambda}{2}  \tilde y} +
  a_1 e^{-i\frac{\eta \lambda}{2} \tilde y}\right]}{I_2 + \gamma_2}.
\end{equation}
The $\rm \gamma_i$ parameters are included not for factorization reasons 
(these wave functions, in quantum cosmology are still non normalizable, except 
when a wave packet is constructed), but as
decoherence parameter embodying a sort of quantum cosmological dissipation
(or damping) distance. The wave function is highly oscillatory, but when the value 
of $\gamma_i$ is increased, the iso-spectral wave function tends to the wave 
function of the original potential (this is similar to the transitory effects 
in quantum mechanics, i.e. probability density in a potential barrier). Unfortunately 
this parameter  $\gamma_i$ can not be used to go from one physical state to another, 
so it is not a time parameter in the usual sense. 
In other words, 
this parameter seems to play the role of a "supersymmetric time-like parameter" 
that gives  the evolution
from a supersymmetric state to the original state, and the usual  behavior for  the wave
function is reached (see figures 3 and 5).  

In the plots of the potentials, we see that the difference between the original 
potential and the isopotential are small at large scale (figure 1), 
but the corresponding wavefunctions present a drastically different behavior to a 
point that it  vanishes  for the limit $\gamma_i \to 0$ and recovers the 
original shape when $\gamma_i \to \infty$. But in all cases, the roots of  the 
wave functions remain the same. This can be attributed to the different structure 
of the iso-spectral potential, as it appears in figures 2 and 3, in the sense that the 
amplitude of the wavefunctions corresponding to the iso-potentials in the regime 
$x<0$, is reduced dramatically.

One expects that these effects should appear in the dynamics of our universe, in 
particular in the inflationary epoch. For inflation we need material with the 
unusual property of negative pressure (i. e. scalar fields), as already mentioned, 
the potential $V(\phi)=e^{-\lambda\phi}$ is one of the most studied. The standard 
approximation technique for analyzing inflation is the slow-roll approximation  
given by the slow roll parameter $\epsilon=\frac{1}{2}(\frac{V'(\phi)}{V(\phi)})^2<<1$ 
in units of $M_p=1$, in particular for the exponential potential 
$\epsilon=\frac{1}{2}\lambda^2<<1$, this means that once we fix $\lambda$ to have 
inflation, there is no mechanism to end it (we can not violate the slow roll 
condition). For the iso-spectral potentials we can see from the plot that we have 
an oscillatory function superposed to the original potential and this oscillatory 
behavior leads to mechanism that ends inflation.
 To see this, we start by writing the slow roll parameter in terms of the potential 
 $V(x)$, it takes the form  $\epsilon=\frac{1}{2}a^2(\frac{V'(x)}{V(x)}+b)^2$, where
 $a=\frac{36-\lambda^2}{\lambda}$ and $b=\frac{36}{\lambda^2-6^2}$, for the original
 potential $\frac{V'(x)}{V(x)}=1$. So again, once we fix the value of $\lambda$ to 
 have inflation there is no way to end it. For the iso-spectral potentials the
  approximate relation holds 
  $\frac{V_{iso}(\phi)}{V(\phi)}{\approx}\frac{V_{iso}(x)}{V(x)}$ where 
  $V_{iso}(x)$ is the iso-spectral potential, and again we may write the 
  slow roll parameter in a similar way 
  $\epsilon=\frac{1}{2}a^2(\frac{V'_{iso}(x)}{V_{iso}(x)}+b)^2$, where the constants 
  $a$ and $b$ depend at most on $\lambda$, now in contrast with the original 
  potential, once we fix the value of $\lambda$ for inflation, the ratio 
  $\frac{V'_{iso}(x)}{V_{iso}(x)}$ can be as large as we want.  We can see 
  this from the plots of the iso-potentials. If $\gamma_i$ is small, the potential 
  is oscillatory and has several roots. In particular near one of the roots 
  $x_i$ $V'_{iso}\ne 0$, but $V(x)\to0$ as $x\to x_i$, that is, in the 
  neighborhood of $x_i$, $\frac{V_{iso}(x)}{V(x)}$ is large, violating the 
  slow-roll conditions, this gives the mechanism to end inflation!.
 Moreover, once inflation has ended the iso -potentials have several minima, 
 where the scalar field can oscillate and start the reheating process. Finally 
 there is a non zero probability that the scalar field may tunnel out of this 
 minimum and give start another inflationary epoch.
\begin{figure}
\includegraphics[width=8cm]{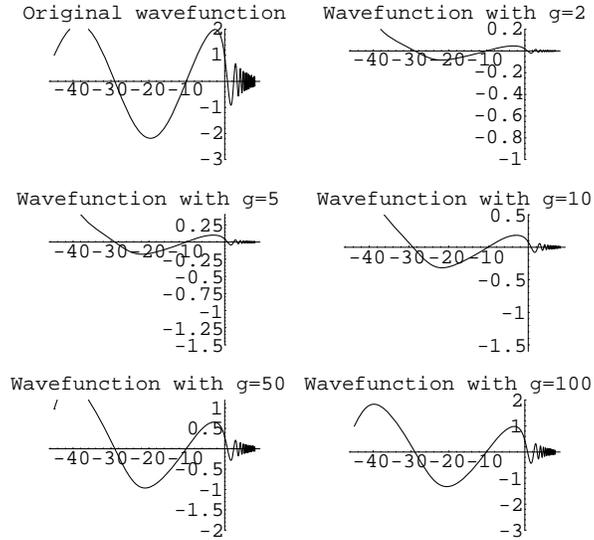}
\caption{\label{fig:esc4} The iso-wave function for different values for the $\gamma$
parameter. The behavior for small values of  $\gamma$ can be seen. This 
different behavior can have effects in the quantum perturbations, and may 
modify structure formation in the large scale universe.}
\end{figure}

\begin{figure}
\includegraphics[width=8cm]{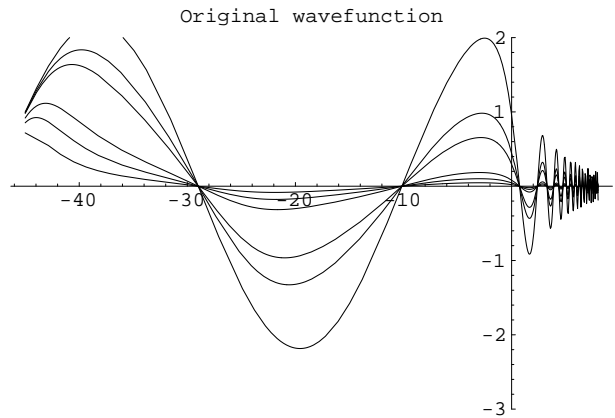}
\caption{\label{fig:esc5} This plot show how the iso-wave functions  tend 
to original wave function
when the parameter $\gamma \to \infty$. }
\end{figure}
\section{Conclusions and Outlook}
In this paper we have applied the factorization method to inflationary cosmology, 
we started with a scalar field model coupled to and FRW model and an exponential 
potential $V(\phi)=V_0e^{-\lambda\phi}$. We have shown that in the variables 
$x$ and $y$ we can reproduce the known result, that for $\lambda<2$ and get
 power law inflation. This new set of variables is important to apply the 
 factorization approach of SUSY-QM, from which we have obtained a complete 
 family of iso-spectral potentials and their corresponding wave functions. 
 We have shown in the plots, that we obtain the original potential and wave 
 function in the limit $\gamma_i\to \infty$. Out of this limit we have a 
 large class of potentials that have the same quantum spectrum (this can 
 clearly be seen in Fig. 5). Therefore  all of these potentials describe 
 the same quantum system. These iso-potentials have a very similar behavior 
 for large scales, so at the classical level we won't see different dynamics. 
 The drastic differences in the iso-wave functions and the iso-potentials, have 
 an impact on the the dynamics of the universe. We have seen that this family 
 of iso-potentials, have a natural mechanism to end inflation, and present the 
 possibility of new inflationary epochs. 
We have found that the $\gamma_i$ parameter plays the role of a decoherence 
parameter that embodies a type of quantum cosmological damping distance. In 
other words, it plays the role of a supersymmetric time-like parameter.
This procedure can be applied to other models like, tachyon driven cosmology 
( Garcia-Compean, Garc\'{\i}a-Jimenez, Obreg\'on and Ram\'{\i}rez, 2005), 
or to toy models of the landscape of string theory 
(Kobakhidze and  Mersini-Houghton, 2004). These results may point to the existence
of other potentials with the same eigenvalue spectrum, so that in principle the 
landscape may be falsifiable. These ideas are being explored and will be presented 
elsewhere.

\acknowledgments{ \noindent This work was partially supported by CONACYT grants 42748 and
 47641, PROMEP grant UGTO-CA-3. MS is also supported by  PROMEP-PTC-085.}

\noindent References

Bagchi B. K. (2001). {\it Supersymmetry in Quantum and Classical Mechanics},
Chapman \& Hall, New York.

Bagrov B.G. and  Samsonov B.F. (1995), {\it Theoretical Mathematical  Physics}. {\bf 104},
356.

Cooper F, Khare A and Sukhatme U. (1995). {\it Physics Repert} {\bf 251}, 267.

Chimento L.P. and Jakubi A. S. (1996). {\it International Journal Modern Physics D} {\bf 5},
 71.
 
Darboux G (1882). {\it C.R. Acad. Sci. (Paris)} {\bf 94}, 1456.

Douglas M. R. (2003). {\it JHEP} {\bf 305}, 046.

Fern\'andez D. J. (1984). {\it Letters Mathematatical Physics} {\bf 8}, 337.

Garcia-Compean H,  Garc\'{\i}a-Jimenez G, Obreg\'on O
	and Ram\'{\i}rez C. (2005)  {\it Physical Review D} {\bf 71}, 063517.
	
Gelfand I. M. and  Levitan B. M. (1955). {\it American Mathematical Society Transl.} {bf 1},
253.
	
Gorini V, Kamenshchik A.Y, Moschella U and Pasquier V. (2004)
    {\it Physical Review D} {\bf 69}, 123512.
    
Gradshteyn I. S and  Ryzhik I. M. (1980). {\it Table of Integrals, Series,
and Products}, (Academic Press, 1980), pages 773-774.   

Guzm\'an W, Sabido M, Socorro J. and Arturo Urena L\'opez L. 
{\it Scalar potential out of canonical quantum cosmology}, gr-qc/0506041.

Ince E. L. (1926) {\it  Ordinary Differential Equations}, Dover, New York.

Junker G. (1996). {\it Supersymmetric Methods in Quantum and Statistical Physics},
Springer, Berlin.
 
Kiefer C. (1988). {\it Physical Review D} {\bf 38}, 1761.
 
Kiefer C. (1990). {\it Nuclear  Physics B} {\bf 341}, 273.

Kobakhidze A and  Mersini-Houghton L. hep-th/04\-10\-213. 

Mersini-Houghton L.  (2005). {\it Classical and  Quantum Gravity} {\bf 22}:3481-3490.

Mielnik B. (1984). {\it  Journal Mathematical Physics} {\bf 25}, 3387.

Nieto M. M. (1984). {\it Physics Letters B} {\bf 145}, 208.

Pappademos J, Sukhatme U. and Pagnamenta A. (1993). {\it Physical Review A} {\bf 48}, 3525.

Rosu H. and Socorro J. (1996). {\it Physics Letters A} {\bf 223}, 28.

Rosu H. and Socorro J. (1998). {\it Il Nuovo Cimento B} {\bf 113}, 683.

Ryan  M Jr. (1972). {\it Hamiltonian Cosmology}, (Springer Verlag).

Samsonov B. F. and Suzko A. A. {\it Discrete supersymmetries
of the Schr\"odinger equation and non-local exactly solvable potential},
quant-ph/0301109.

Sen A. (2002). {\it Modern Physics Letters A} {\bf 17}, 1797.
 
Sen A. (2003). {\it International Journal Modern Physics A} {\bf 18}, 4869.

Socorro J. Reyes M. A. and  Gelbert F.A. (2003). {\it Physics Letters A} {\bf 313}, 338-342.

Susskind L. hep-th/0302219.

\end{document}